# Counterexample-Preserving Reduction for Symbolic Model Checking

Wanwei Liu, Rui Wang, Xianjin Fu,
Ji Wang, Wei Dong, and Xiaoguang Mao

School of Computer Science,
National University of Defense Technology,
Changsha, P.R. China, 410073

**Abstract.** The cost of LTL model checking is highly sensitive to the length of the formula under verification. We observe that, under some specific conditions, the input LTL formula can be reduced to an easier-to-handle one before model checking. In our reduction, these two formulae need not to be logically equivalent, but they share the same counterexample set w.r.t the model. In the case that the model is symbolically represented, the condition enabling such reduction can be detected with a lightweight effort (e.g., with SAT-solving). In this paper, we tentatively name such technique "CounterExample-Preserving REduction" (CePRe, for short), and finally the proposed technqiue is experimentally evaluated by adapting NuSMV.

## 1 Introduction

Linear Temporal Logic (LTL, for short) [11] is one of the most frequently used specification languages in model checking (cf. [14]). It designates properties over a linear structure, which can be viewed as an execution of the program. The task of LTL model checking is to search the state space (explicitly or implicitly), with the goal of detecting the existence of feasible traces violating the specification. If such traces exist, the model checker will report one of them as a "counterexample"; otherwise, the model checker will give an affirmative report.

It can be shown that the complexity of LTL model checking $M \models \varphi$ is in $\mathcal{O}(|M| \times 2^{|\varphi|})$, meanwhile, the nesting depth of temporal operators might be the major factor affecting the cost in compiling LTL formulae.

Hence, it is reasonable to simplify the specification before conducting model checking. For example, in [12], Somenzi and Bloem provided a series of rewriting schemas for simplifying LTL specifications, and these rewriting schamas preserve logical equivalence.

One may argue that "a majority of LTL formulas used in real applications are simple, succinct rather than complicated", but, we need to notice the following facts:

- Typically, the LTL formula $\mathbf{F}(p\mathbf{U}q)$ is usually considered as a "simple" one, nevertheless, it can be further simplified to $\mathbf{F}q$, and this fact tends to be omitted.[1]
- Indeed, people do use complicate specifications in the real industrial field, as well in some standard benchmark (cf. [2]).
- Last but not least, not all specifications are designated manually. Actually, some formulae are generated by specification-generaton-tools (e.g., PROSPEC). Indeed, one may find that lots of these machine-generated specifications can be simplified.

Symbolic model checking [10] is one of the most significant breakthrough in model checking, and two major fashions of symbolic model checking are widely used: one is the BDD-based manner [6], and the other is SAT-based manner, such as bounded model checking [1].

Instead of using an explicit representation, the symbolic manner represents state space with a series of Boolean formulae. This enables implicit manipulation of the verification process and it usually leads to an efficient implementation [3]. Meanwhile, such a unified representation of transitions and invariants of the model potentially provides heuristic information to simplify the specification. For example:

- The formulae $p\mathbf{U}q$ and $(r\mathbf{U}p)\mathbf{U}q$ can be respectively reduced as $q$ and $(r\mathbf{U}p) \vee q$, if we know that $p \rightarrow q$ holds everywhere in the model.
- Each occurrence of $\mathbf{G}\theta$ in the specification can be replaced with $\top$ (i.e., logically true), if we can inductively infer that the Boolean formula $\theta$ holds at each reachable state in the model.

Actually, we can make certain of these conditions with the following efforts.

- To check whether "$p \rightarrow q$ holds everywhere in the model", we may test if $p \rightarrow q$ is an *invariant* in the model — i.e., if $\rho \wedge \neg(p \rightarrow q)$ is unsatisfiable (we in the later denote it as $\rho \vdash p \rightarrow q$), where $\rho$ is the Boolean encoding of the model's transition relation.
- Likely, to justify that $\theta$ holds at each reachable state, it suffices to ensure that $\theta_0 \vdash \theta$ and $\rho \vdash \theta \rightarrow \theta'$, where $\theta_0$ is the initial condition of the model.

Hence, this provides an opportunity to replace the specification with a simpler one, accompanied with some lightweight extra task of condition detection. Even if such detection fails, the overhead is usually negligible.

In this paper, we systematically investigate the above idea, and tentatively name this technique *CounterExample-Preserving REduction* (CEPRE, for short). Such reduction can be done before starting model checking, and it is an orthogonal optimization technique to both encoding approaches and model compression techniques.

To justify it, we have extended NuSMV and implement CEPRE as an upfront option for LTL model checking. Subsequently, we conduct experiments over both industrial benchmarks and randomly generated cases. Experimental results show that CePRe can improve the efficiency significantly.

---

[1] On one hand, $p\mathbf{U}q$ implies $\mathbf{F}q$, and hence $\mathbf{F}(p\mathbf{U}q)$ implies $\mathbf{F}\mathbf{F}q$ (i.e., $\mathbf{F}q$); on the other hand, $q$ implies $p\mathbf{U}q$, and hence $\mathbf{F}q$ implies $\mathbf{F}(p\mathbf{U}q)$.

This paper is organized as follows: Section 2 revisits some basic notions. Section 3 introduces the CePRe technique and gives the performance analysis. In Section 4, experimental results over industrial benchmarks as well over random generated cases are given. We summarize the whole paper with Section 5.

## 2 Preliminaries

We presuppose a countable set $\mathcal{P}$ of *atomic propositions*, ranging over $p$, $q$, $p_1$, etc. For each proposition $p \in \mathcal{P}$, we create a *primed version* $p'$ (not belonging to $\mathcal{P}$) for it. For each set $\mathcal{V} \subseteq \mathcal{P}$, we define $\mathcal{V}' \triangleq \{p' \mid p \in \mathcal{V}\}$. We use $\mathbf{B}(\mathcal{V})$ to denote the set of Boolean formulae over $\mathcal{V}$, similarly, we denote by $\mathbf{B}(\mathcal{V} \cup \mathcal{V}')$ the set of Boolean formulae over $\mathcal{V} \cup \mathcal{V}'$. The scope of the *prime* operator can be naturally lifted to Boolean formulae over $\mathbf{B}(\mathcal{V})$, by defining

$$(\neg \theta)' \triangleq \neg \theta' \qquad (\theta_1 \wedge \theta_2)' \triangleq \theta_1' \wedge \theta_2' \quad (\theta_1 \vee \theta_2)' \triangleq \theta_1' \vee \theta_2'.$$

An *assignment* is a subset $\mathcal{V}$ of $\mathcal{P}$, intuitively, it assigns 1 (or, true) to propositions belonging to $\mathcal{V}$, and assigns 0 (or, false) to other propositions. For each $\mathcal{V} \subseteq \mathcal{U} \subseteq \mathcal{P}$ and $\theta \in \mathbf{B}(\mathcal{U})$, we denote by $\mathcal{V} \Vdash \theta$ if $\theta$ is evaluated to 1 under the assignment $\mathcal{V}$.

A *united assignment* is a pair $(\mathcal{V}_1, \mathcal{V}_2)$, where both $\mathcal{V}_1$ and $\mathcal{V}_2$ are subsets of $\mathcal{P}$. It assigns 1 to propositions belonging to $\mathcal{V}_1 \cup \mathcal{V}_2'$, and assigns 0 to other propositions. Suppose that $\mathcal{V}_1, \mathcal{V}_2 \subseteq \mathcal{U} \subseteq \mathcal{P}$ and $\theta \in \mathbf{B}(\mathcal{U} \cup \mathcal{U}')$, we also write $(\mathcal{V}_1, \mathcal{V}_2) \Vdash \theta$ if $\theta$ is evaluated to 1 under the united assignment $(\mathcal{V}_1, \mathcal{V}_2)$.

LTL formulae can be inductively defined as follows.

- $\bot$ and $\top$ are LTL formulae.
- Each proposition $p \in \mathcal{P}$ is an LTL formula.
- If both $\varphi_1$ and $\varphi_2$ are LTL formulae, so does $\varphi_1 \to \varphi_2$.
- If $\varphi$ is an LTL formula, then $\mathbf{X}\varphi$ and $\mathbf{Y}\varphi$ are LTL formulae.
- If $\varphi_1$ and $\varphi_2$ are LTL formulae, then both $\varphi_1 \mathbf{U} \varphi_2$ and $\varphi_1 \mathbf{S} \varphi_2$ are LTL formulae.

Semantics of an LTL formula is defined w.r.t. a *linear structure* $\pi \in (2^{\mathcal{P}})^\omega$ (i.e., $\pi$ is an infinite word over the alphabet $2^{\mathcal{P}}$) and a position $i \prec \omega$. Inductively:

- $\pi, i \models \top$ and $\pi, i \not\models \bot$;
- $\pi, i \models p$ iff $\pi(i) \Vdash p$ (where $\pi(i)$ is the $i$-th letter of $\pi$, which can be viewed as an assignment for it is a subset of $\mathcal{P}$);
- $\pi, i \models \varphi_1 \to \varphi_2$ iff either $\pi, i \not\models \varphi_1$ or $\pi, i \models \varphi_2$;
- $\pi, i \models \mathbf{X}\varphi$ iff $\pi, i+1 \models \varphi$;
- $\pi, i \models \mathbf{Y}\varphi$ iff $i > 0$ and $\pi, i-1 \models \varphi$;
- $\pi, i \models \varphi_1 \mathbf{U} \varphi_2$ iff there is some $j \geq i$, s.t. $\pi, j \models \varphi_2$ and $\pi, k \models \varphi_1$ for each $i \leq k < j$;

- $\pi, i \models \varphi_1 \mathbf{S} \varphi_2$ iff there is some $j \leq i$, s.t. $\pi, j \models \varphi_2$ and $\pi, k \models \varphi_1$ for each $i \geq k > j$.

For the sake of convenience, we usually directly write $\pi, 0 \models \varphi$ as $\pi \models \varphi$.

As usual, we employ some derived Boolean connectives such as

$$\neg \varphi \triangleq \varphi \to \bot \qquad \varphi \vee \psi \triangleq \neg \varphi \to \psi \qquad \varphi \wedge \psi \triangleq \neg(\neg \varphi \vee \neg \psi)$$

and derived temporal operators such as

$$\begin{aligned} \mathbf{F}\varphi &\triangleq \top \mathbf{U}\varphi & \mathbf{Z}\varphi &\triangleq \neg \mathbf{Y} \neg \varphi & \mathbf{O}\varphi &\triangleq \top \mathbf{S}\varphi \\ \mathbf{G}\varphi &\triangleq \neg \mathbf{F} \neg \varphi & & & \mathbf{H}\varphi &\triangleq \neg \mathbf{O} \neg \varphi \\ \varphi \mathbf{R} \psi &\triangleq \neg(\neg \varphi \mathbf{U} \neg \psi) & & & \varphi \mathbf{T} \psi &\triangleq \neg(\neg \varphi \mathbf{S} \neg \psi) \end{aligned}$$

We say that $\wedge$ and $\vee$, $\mathbf{F}$ and $\mathbf{G}$, $\mathbf{O}$ and $\mathbf{H}$, $\mathbf{Y}$ and $\mathbf{Z}$, $\mathbf{U}$ and $\mathbf{R}$, $\mathbf{T}$ and $\mathbf{S}$ are pairwise the *dual operators*.

Temporal operators like $\mathbf{X}$, $\mathbf{U}$, $\mathbf{F}$, $\mathbf{G}$, $\mathbf{R}$ are called *future operators*, whereas $\mathbf{Y}$, $\mathbf{Z}$, $\mathbf{S}$, $\mathbf{O}$, $\mathbf{H}$ and $\mathbf{T}$ are called *past operators*. We say an LTL is *pure future* (resp. *pure past*) if it involves no past (resp. future) operators.

**Theorem 1 ([7]).** *Each LTL formula has an equivalent pure future expression.*

Theorem 1 tells the fact that past operators do not add any expressive power to LTL formulae. Nevertheless, with these, we can give a much more succinct description in defining specifications.

Given an LTL formula $\varphi$, we denote by $sub(\varphi)$ the set constituted with subformulae of $\varphi$. Particularly, we respectively denote by $sub_\mathbf{U}(\varphi)$ and $sub_\mathbf{S}(\varphi)$ the set of $\varphi$'s subformulae consisting of "$\mathbf{U}$-subformulae" and "$\mathbf{S}$-subfomulae". An $\mathbf{U}$-formula (resp. $\mathbf{S}$-formula) is a formula rooted at $\mathbf{U}$ (resp. $\mathbf{S}$).

A *model* is a tuple $M = \langle \mathcal{V}, \rho, \theta_0, \mathcal{F}, \mathcal{C} \rangle$, where

- $\mathcal{V} \subseteq \mathcal{P}$, is a finite set of atomic propositions.
- $\rho \in \mathbf{B}(\mathcal{V} \cup \mathcal{V}')$, is the *transition relation*.
- $\theta_0 \in \mathbf{B}(\mathcal{V})$, is the *initial condition*.
- $\mathcal{F} \subseteq \mathbf{B}(\mathcal{V})$, is a set of *fairness constraints*.
- $\mathcal{C} \subseteq \mathbf{B}(\mathcal{V}) \times \mathbf{B}(\mathcal{V})$, is a set of *compassion constraints*.

A *derived linear structure* of $M$ is an infinite word $\pi \in (2^\mathcal{V})^\omega$, such that

1. $\pi(0) \Vdash \theta_0$;
2. $(\pi(i), \pi(i+1)) \Vdash \rho$ for each $i \prec \omega$;
3. for each $\varphi \in \mathcal{F}$, there are infinitely many $i$'s having $\pi(i) \Vdash \varphi$;
4. for each $(\varphi, \psi) \in \mathcal{C}$, if there are infinitely many $i$'s having $\pi(i) \Vdash \varphi$, then there are also infinitely many $j$'s such that $\pi(j) \Vdash \psi$.

We denote by $\mathbf{L}(M)$ the set of derived linear strctures of $M$, call it the *language* of $M$.

For a model $M$ and an LTL formula $\varphi$, we denote as $M \models \varphi$ if $\pi \models \varphi$ for each $\pi \in \mathbf{L}(M)$. Meanwhile, we define

$$\mathbf{CE}(\varphi, M) \triangleq \{\pi \in \mathbf{L}(M) \mid \pi \not\models \varphi\}$$

and call it the *counterexample set* of $\varphi$ w.r.t. $M$.

# 3 Counterexample-Preserving Reduction

We describe the CEPRE technique in this section, and we would fix components of the model $M$, and just let it be $\langle \mathcal{V}, \rho, \theta_0, \mathcal{F}, \mathcal{C} \rangle$.

For $M$, we are particularly concerned about formulae having the same counterexample set — we say that $\varphi$ and $\psi$ are *inter-reduce-able* w.r.t. $M$ if and only if $\mathbf{CE}(\varphi, M) = \mathbf{CE}(\psi, M)$, denoted as $\varphi \approx_M \psi$. Hence, $\varphi \approx_M \psi$ implies that $M \models \varphi \Leftrightarrow M \models \psi$.

The central part of CEPRE is a series of *reduction rules* being of the form

$$\text{Cond} \ \triangleright \ \varphi \approx_M \psi \quad (\text{NAME})$$

where "Cond" is called the *additional condition*.

Though the relation $\approx_M$ is, actually symmetric, we always write the formula being reduced on the righthand of the "$\approx$" sign in reduction rules. Since the model $M$ is fixed, in this section, we omit it from the subscript. In addition, if the additional condition trivially holds, we will discard this part, and directly write the rule as $\varphi \approx \psi$, and we say such a reduction rule "*model-independent*"; in contrast, we call other rules "*model-dependent*".

## 3.1 The Reduction Rules

First of all, we have some elementary reduction rules as depicted in Figure 1. For the rules (INIT), (IND) and (TRANS), the notation "$\vdash$" occurring in the condition part standards "*inferring*" relation in propositional logic ($\rho \vdash \theta$ iff $\rho \wedge \neg \theta$ is unsatisfiable), and we here require that $\theta, \theta_1, \theta_2 \in \mathbf{B}(\mathcal{V})$.

---

$\theta_0 \vdash \theta \ \triangleright \ \theta \approx \top \quad (\text{INIT}) \qquad\qquad \rho \vdash \theta \ \triangleright \ \mathbf{G}\theta \approx \top \quad (\text{TRANS})$

$\theta_0 \vdash \theta; \ \rho \vdash \theta \to \theta' \ \triangleright \ \mathbf{G}\theta \approx \top \quad (\text{IND})$

$\theta \in \mathcal{F} \ \triangleright \ \mathbf{GF}\theta \approx \top \quad (\text{FAIR}) \qquad (\theta_1, \theta_2) \in \mathcal{C} \ \triangleright \ (\mathbf{GF}\theta_1 \to \mathbf{GF}\theta_2) \approx \top \quad (\text{COMP})$

---

**Fig. 1.** Elementary reduction rules.

Subsequently, let us define a partial order "$\sqsubseteq$" over unary temporal operators (and their combinations) as follows:

$$\mathbf{F} \sqsubseteq \mathbf{GF} \sqsubseteq \mathbf{FG} \sqsubseteq \mathbf{G}$$
$$\mathbf{F} \sqsubseteq \mathbf{X}^i \sqsubseteq \mathbf{G} \quad (i \prec \omega)$$
$$\mathbf{O} \sqsubseteq \mathbf{HO} \sqsubseteq \mathbf{OH} \sqsubseteq \mathbf{H}$$

where $\mathbf{X}^0 \varphi \triangleq \varphi$ and $\mathbf{X}^{i+1} \varphi \triangleq \mathbf{X}(\mathbf{X}^i \varphi)$.

Assume that $\mathbf{P}^w, \mathbf{P}^s \in \{\mathbf{F}, \mathbf{FG}, \mathbf{GF}, \mathbf{G}, \mathbf{O}, \mathbf{HO}, \mathbf{OH}, \mathbf{H}\} \cup \{\mathbf{X}^i \mid i \prec \omega\}$ and $\mathbf{P}^w \sqsubseteq \mathbf{P}^s$, then we have two model-indenpendent rules, as depicted in Figure 2. Though these rules seem to be trivial, they are useful in doing combinational reductions (see the example given in Section 3.2).

$$(\mathbf{P}^w\varphi \wedge \mathbf{P}^s\varphi) \approx \mathbf{P}^s\varphi \quad \text{(Conj)} \qquad (\mathbf{P}^w\varphi \vee \mathbf{P}^s\varphi) \approx \mathbf{P}^w\varphi \quad \text{(Disj)}$$

**Fig. 2.** Reduction rules of (Conj) and (Disj).

Figure 3 provides some reduction rules that can be used to simplify nested temporal operators. Moreover, we may immediately get such a rule's "past version" by switching $\mathbf{U}$ and $\mathbf{S}$, $\mathbf{R}$ and $\mathbf{T}$, etc. For example, we may obtain the rule (OS) (i.e., $\mathbf{O}(\varphi \mathbf{S} \psi) \approx \mathbf{O}\psi$) from (FU) .

$$\mathbf{F}(\varphi \mathbf{U} \psi) \approx \mathbf{F}\psi \quad \text{(FU)} \qquad \varphi\mathbf{U}(\mathbf{F}\psi) \approx \mathbf{F}\psi \quad (\text{U}_\text{F})$$

$$\mathbf{FF}\varphi \approx \mathbf{F}\varphi \quad \text{(FF)} \qquad \mathbf{GFG}\varphi \approx \mathbf{FG}\varphi \quad \text{(GFG)}$$

**Fig. 3.** Reduction rules for formulae involving nested pure future operators.

Meanwhile, we also have the *Duality Principle* for model-independent rules: "by switching each operator with its dual operator, then we may get a new reduction rule". For the rules listed in Figure 3, we may obtain the corresponding rules such as (GR), (R$_\text{G}$), (GG) and (FGF). As an example, the rule (GG) is just $\mathbf{GG}\varphi \approx \mathbf{G}\varphi$.

$$\mathbf{Y}\varphi \approx \bot \quad \text{(Y)} \qquad \mathbf{O}\varphi \approx \varphi \quad \text{(O)} \qquad \varphi\mathbf{S}\psi \approx \varphi \quad \text{(S)}$$

**Fig. 4.** Reduction rules for formulae involving (outermost) past operators.

Since we always stand at the starting point when doing model checking (i.e., the goal is to check if $\pi, 0 \models \varphi$ for each $\pi \in \mathbf{L}(M)$), hence, we can sometimes "erase" the outermost past operators, as shown in Figure 4. Note that we can also acquire the rules (Z), (H) and (T) according to the Duality Principle. Just beware the exception that the rule (Z) should be $\mathbf{Z}\varphi \approx \top$.

---

$$\mathbf{XY}\varphi \approx \varphi \quad \text{(XY)} \qquad \mathbf{FH}\varphi \approx \mathbf{H}\varphi \quad \text{(FH)}$$

$$\mathbf{FO}\varphi \approx \mathbf{F}\varphi \vee \mathbf{O}\varphi \quad \text{(FO)} \qquad \mathbf{F}(\varphi\mathbf{S}\psi) \approx \mathbf{F}\psi \vee \varphi\mathbf{S}\psi \quad \text{(FS)}$$

---

**Fig. 5.** Reduction rules for formulae involving adjacent past and future operators.

Figure 5 introduces a series of rules handing formulae involving adjacent past and future temporal operators. Remind that the rules (XZ), (GO), (GH) and (GT) are also immediately available.

---

$$\rho \vdash \theta_1 \vee \theta_2 \triangleright \theta_1\mathbf{U}\theta_2 \approx \mathbf{F}\theta_2 \quad \text{(U)}$$

$$\rho \vdash \theta_2 \to \theta_1 \vee \theta_2' \triangleright \theta_1\mathbf{R}\theta_2 \approx \theta_2 \quad \text{(R)}$$

---

**Fig. 6.** Reduction rules of (U) and (R).

From now on, we let $\theta_1, \theta_2, \ldots$ range over $\mathbf{B}(\mathcal{V})$, and let $\varphi_1, \varphi_2, \ldots$ be arbitrary LTL formulae. We have some model-dependent rules. The first group of such rules are listed in Figure 6.

In Figure 7, another set of reduction rules are provided, and these rules are mainly concerned with LTL formulae involving adjacent $\mathbf{U}$-operators. Note that when applying the Duality principle to model dependent rules, besides switching the operators, we also need to exchange the antecedent and subsequent in the condition part. As an example, we may obtain the reduction rule

$$\rho \vdash \theta_3 \to \theta_2 \triangleright (\varphi_1\mathbf{R}\theta_2)\mathbf{R}\theta_3 \approx \theta_3 \wedge (\varphi_1\mathbf{R}\theta_2) \quad (\mathrm{R}^{\mathrm{R}}[3 \to 2])$$

by applying the Duality Principle to $(\mathrm{U}^{\mathrm{U}}[2 \to 3])$.

Lastly, Figure 8 provides some reduction rules that can be used to simplify formulae with mixed usage of $\mathbf{U}$ and $\mathbf{R}$. Similarly, dualize operators and inverse the additional condition, one may obtain reduction rules for formulae in which $\mathbf{R}$ appears (adjancently) out of $\mathbf{U}$.

### 3.2 Reduction Strategy

We show the usage of CePRe reduction rules by illustrating the reduction process of $M \models (\theta_1\mathbf{U}\theta_2)\mathbf{U}\theta_3$:

1. We may first try with the rule $(\mathrm{U}^{\mathrm{U}}[1 \to 3])$ by inquiring the SAT-solver if $\rho \vdash \theta_1 \to \theta_3$ holds.

$$\rho \vdash \theta_1 \to \theta_2 \rhd \theta_1 \mathbf{U}\theta_2 \approx \theta_2 \qquad (\mathrm{U}[1 \to 2])$$

$$\rho \vdash \theta_1 \to \theta_3 \rhd (\theta_1 \mathbf{U}\varphi_2)\mathbf{U}\theta_3 \approx \varphi_2 \mathbf{U}\theta_3 \qquad (\mathrm{U}^{\mathrm{U}}[1 \to 3])$$

$$\rho \vdash \theta_2 \to \theta_3 \rhd (\varphi_1 \mathbf{U}\theta_2)\mathbf{U}\theta_3 \approx \theta_3 \vee (\varphi_1 \mathbf{U}\theta_2) \qquad (\mathrm{U}^{\mathrm{U}}[2 \to 3])$$

$$\rho \vdash \theta_3 \to \theta_2 \rhd (\varphi_1 \mathbf{U}\theta_2)\mathbf{U}\theta_3 \approx (\varphi_1 \vee \theta_2)\mathbf{U}\theta_3 \qquad (\mathrm{U}^{\mathrm{U}}[3 \to 2])$$

$$\rho \vdash \theta_2 \to \theta_3' \rhd (\varphi_1 \mathbf{U}\theta_2)\mathbf{U}\theta_3 \approx (\varphi_1 \vee \theta_2)\mathbf{U}\theta_3 \qquad (\mathrm{U}^{\mathrm{U}}[2 \to 3'])$$

$$\rho \vdash \neg\theta_2 \to \theta_3 \rhd (\varphi_1 \mathbf{U}\theta_2)\mathbf{U}\theta_3 \approx \mathbf{F}\theta_3 \qquad (\mathrm{U}^{\mathrm{U}}[\neg 2 \to 3])$$

$$\rho \vdash \theta_1 \to \theta_2 \rhd \theta_1 \mathbf{U}(\theta_2 \mathbf{U}\varphi_3) \approx \theta_2 \mathbf{U}\varphi_3 \qquad (\mathrm{U}_{\mathrm{U}}[1 \to 2])$$

$$\rho \vdash \theta_1 \to \theta_3 \rhd \theta_1 \mathbf{U}(\varphi_2 \mathbf{U}\theta_3) \approx \varphi_2 \mathbf{U}\theta_3 \qquad (\mathrm{U}_{\mathrm{U}}[1 \to 3])$$

$$\rho \vdash \theta_2 \to \theta_1 \rhd \theta_1 \mathbf{U}(\theta_2 \mathbf{U}\varphi_3) \approx \theta_1 \mathbf{U}\varphi_3 \qquad (\mathrm{U}_{\mathrm{U}}[2 \to 1])$$

**Fig. 7.** Reduction rules for formulae involving adjacent **U** operators.

$$\rho \vdash \theta_1 \to \theta_3 \rhd (\theta_1 \mathbf{R}\varphi_2)\mathbf{U}\theta_3 \approx ((\theta_1 \mathbf{R}\varphi_2) \vee \theta_3) \wedge \mathbf{F}\theta_3 \qquad (\mathrm{U}^{\mathrm{R}}[1 \to 3])$$

$$\rho \vdash \neg\theta_1 \to \theta_3 \rhd (\theta_1 \mathbf{R}\varphi_2)\mathbf{U}\theta_3 \approx \varphi_2 \mathbf{U}\theta_3 \qquad (\mathrm{U}^{\mathrm{R}}[\neg 1 \to 3])$$

$$\rho \vdash \theta_1 \to \theta_3 \rhd \theta_1 \mathbf{U}(\varphi_2 \mathbf{R}\theta_3) \approx \varphi_2 \mathbf{R}\theta_3 \qquad (\mathrm{U}_{\mathrm{R}}[1 \to 3])$$

**Fig. 8.** Reduction rules for formulae involving adjacent **U** and **R** operators.

2. If the SAT-solver returns "unsatisfiable" with the input $\rho \wedge \theta_1 \wedge \neg\theta_3$, it implies that the additional condition is stated, and we may replace the specification with $\theta_2 \mathbf{U}\theta_3$.
3. Otherwise, we will try with the next reduction rule, such as ($\mathrm{U}^{\mathrm{U}}[2 \to 3]$).

In fact, these rules can also be "*locally applied*" to subformulae. For example, to make a local reduction of (FU), we may replace each occurrence of $\mathbf{F}(\varphi \mathbf{U}\psi)$ in the specification with $\mathbf{F}\psi$. The only exception is for the group of rules listed in Figure 4: observe that we have $\mathbf{Y}\varphi \approx \bot$ according to (Y), yet this does not imply that $\mathbf{FY}\varphi \approx \mathbf{F}\bot$ holds. Hence, these rules have an "implicit condition" when doing local application: the subformula to be reduced must occur "*temporally outermost*" in the specification — i.e., the target subformula is not in the scope of any temporal operators in the specification.

Compositional use of reduction rules may lead to a more aggressive reduction. For example:

**Input**: The original specification $\varphi$.
**Output**: The specification having been reduced.

```
 1 let Γ := ∅; /* Γ memorizes the sub-formulae with infeasible condition*/
 2 let Δ := {ψ ∈ (sub(φ) \ Γ) such that ψ matches some reduction rule(s)};
 3 foreach ψ₁, ψ₂ ∈ Δ s.t. ψ₁ ≠ ψ₂ do
 4     if ψ₁ ∈ sub(ψ₂) then
 5         Δ := Δ \ {ψ₁}; /* i.e., we only proceed "max" subformulae */
 6     end
 7 end
 8 if Δ = ∅ then
 9     return φ;
10 end
11 foreach ψ ∈ Δ do
12     let Θ := the set of rules that can be applied to ψ;
13         /* note that we have |Θ| ≤ 5 for each ψ */
14     while Θ ≠ ∅ do
15         choose R := (Cond ▷ ψ ≈ η) in Θ ;
16         if Cond is stated then
17             φ := φ^ψ_η;  /* φ^ψ_η is obtained from φ by replacing ψ with η */
18             break;
19         end
20         Θ := Θ \ {R};
21     end
22     Δ := Δ \ {ψ};
23     if Θ = ∅ then
24         Γ := Γ ∪ {ψ} ; /* ψ would be excluded in the next iteration */
25     end
26 end
27 goto 2;
```

**Algorithm 1:** The "max-match" rule-selection strategy.

1. For the task of model checking $M \models \mathbf{FO}p$, we may firstly change the goal as $M \models \mathbf{F}p \vee \mathbf{O}p$, according to the rule (FO).
2. Now, the subformula $\mathbf{O}p$ is a temporally outermost one, hence we may take a local application of (O), and then the goal becomes $M \models \mathbf{F}p \vee p$.
3. Finally, we may change the model checking obligation into $M \models \mathbf{F}p$ via the rule (DISJ).

In the real implementation, we may perform a "*max-match*" rule-selection strategy, as depicted in Algorithm 1. In Line 15, for a rule "Cond $\triangleright\ \psi \approx \eta$",

1. the simpler Cond is, and
2. the shorter $\eta$ is,

the higher priority to be chosen it has. Hence, a model-independent always has a higher priority than a model-dependent one. We can see that the reduction can be accomplished within $\mathcal{O}(|\varphi|)$ iterations.

### 3.3 Performance Analysis of the Reduction

We now try to answer the question "why we can gain a better performance during verification if CePRe is conducted first". To give a rigorous explanation, we briefly revisit the implementation of symbolic model checking algorithms.

The core procedure of BDD-based LTL symbolic model checking algorithm is to construct a *tableau* for the (negated) property. In what followed, we refer the tableau of $\varphi$ as $T_\varphi$, and we would give an analysis on its major components affecting the cost of model checking.

**State space**: The state space of $T_\varphi$ consists of subsets of $el(\varphi)$, and the set $el(\varphi)$ can be inductively computed as follows.

- $el(\top) = el(\bot) = \emptyset$.
- $el(p) = \{p\}$ if $p \in \mathcal{P}$.
- $el(\varphi_1 \to \varphi_2) = el(\varphi_1) \cup el(\varphi_2)$.
- $el(\mathbf{X}\psi) = \{\mathbf{X}\psi\} \cup el(\psi)$, and $el(\mathbf{Y}\psi) = \{\mathbf{Y}\psi\} \cup el(\psi)$.
- $el(\varphi_1 \mathbf{U} \varphi_2) = el(\varphi_1) \cup el(\varphi_2) \cup \{\mathbf{X}(\varphi_1 \mathbf{U} \varphi_2)\}$ and $el(\varphi_1 \mathbf{S} \varphi_2) = el(\varphi_1) \cup el(\varphi_2) \cup \{\mathbf{Y}(\varphi_1 \mathbf{S} \varphi_2)\}$.

With symbolic representation, each formula $\psi \in el(\varphi)$ corresponds to a proposition in building the tableau. Moreover, if $\psi \in \mathcal{P}$, then no new proposition need to be introduced (since it has already been introduced in building the symbolic representation of $M$), otherwise, a fresh proposition $p_\psi$ is required. Hence the total number of newly introduced propositions equals to $|el(\varphi) \setminus \mathcal{P}|$. From an induction over formula's structure, we have the following claim.

**Proposition 1.** *$|el(\varphi) \setminus \mathcal{P}|$ equals to the number of temporal operators in $\varphi$.*

**Transitions**: The transition relation of $T_\varphi$ is a conjunction of a set of constraints, and each constraint is either of the form $p_{\mathbf{X}\psi} \leftrightarrow \sigma'(\psi)$ or $p'_{\mathbf{Y}\eta} \leftrightarrow \sigma(\eta)$, where $\mathbf{X}\psi, \mathbf{Y}\eta \in el(\varphi)$, and the function $\sigma$ can inductively defined as follows.

- $\sigma(\bot) = \bot$ and $\sigma(\top) = \top$.
- $\sigma(p) = p$ for each $p \in \mathcal{P}$.
- $\sigma(\psi_1 \to \psi_2) = \sigma(\psi_1) \to \sigma(\psi_2)$.
- $\sigma(\mathbf{X}\psi_1) = p_{\mathbf{X}\psi_1}$ and $\sigma(\mathbf{Y}\psi_2) = p_{\mathbf{Y}\psi_2}$.
- $\sigma(\psi_1 \mathbf{U} \psi_2) = \sigma(\psi_2) \vee \sigma(\psi_1) \wedge p_{\mathbf{X}(\psi_1 \mathbf{U} \psi_2)}$ and $\sigma(\psi_1 \mathbf{S} \psi_2) = \sigma(\psi_2) \vee \sigma(\psi) \wedge p_{\mathbf{Y}(\psi_1 \mathbf{U} \psi_2)}$.

According to the definition of $el$, we can see that each $\psi \in sub(\varphi)$ rooted at a future (reps. past) temporal operator exactly produces one formula $\mathbf{X}\eta$ (resp. $\mathbf{Y}\eta$) in $el(\varphi)$, and hence a new proposition $p_{\mathbf{X}\eta}$ (resp. $p_{\mathbf{Y}\eta}$) would be introduced. Subsequently, each such $p_{\mathbf{X}\eta}$ (reps. $p_{\mathbf{Y}\eta}$) adds exactly one constraint to the transition relation. Hence, we have the following claim.

**Proposition 2.** *The number of constraints in the transition relation of $T_\varphi$ equals to the number of temporal operators occurring in $\varphi$ (alternatively, $|el(\varphi) \setminus \mathcal{P}|$).*

**Fairness constraints**: According to the tableau construction, each $\psi \in sub_{\mathbf{U}}(\varphi)$ would impose a fairness constraint to $T_\varphi$. Hence, the number of fairness constraints equals to $|sub_{\mathbf{U}}(\varphi)|$.

With a case-by-case checking, we can show the following theorem.

**Theorem 2.** *Let "Cond $\triangleright \varphi \approx \psi$" be a reduction rule, then we have $|el(\psi)\backslash\mathcal{P}| \leq |el(\varphi) \backslash \mathcal{P}|$ and $|sub_{\mathbf{U}}(\psi)| \leq |sub_{\mathbf{U}}(\varphi)|$.*

In contrast, the cost of BMC is quite sensitive to the encoding approach. In a broad sense, we can categorize the encoding approaches into two fashions.

**Syntactic encoding** Such kind of encodings are inductively produced w.r.t. the formula's structure. The very original one is presented in [1], and this is improved in [4] by observing some properties of that encoding. In [9] (as well in [2]), a linear incremental syntactic encoding is suggested.

**Semantic encoding** In [5], an alternative BMC technique is provided: it mimics the tableau-based model checking process, but it express the fair-path detection upon the product model with Boolean formula.[2]

For the semantic encodings, the reason that we can benefit from CEPRE is exactly the same as that for BDD-based approach. Because, the encoding is a conjunction of a $k$-step unrolling of $M$ and a $k$-step unrolling of $T_\varphi$ (an unrolling is either a partial linear structure, or a one ending with a loop). The former is usually in a fixed pattern, and for the latter we need $k \times |el(\varphi) \setminus \mathcal{P}|$ new propositions, and the sizes of Boolean formulae w.r.t the transition and fairness constraints[3] are respectively $\mathcal{O}(k \times |el(\varphi) \setminus \mathcal{P}|)$ and $\mathcal{O}(k^2 \times |sub_{\mathbf{U}}(\varphi)|)$.

For a syntactic BMC encoding, one need to generate a Boolean formula of the form $E_M^k \wedge E_{\neg\varphi}^k$, where $E_M^k$ is the "unrolling" of $M$ with $k$ steps, and $E_{\neg\varphi}^k$ describes that the $k$-step unrolling causes a violation of $\varphi$. In general, $E_M^k$ is almost the same in all syntactic encodings, and the key factor affecting the cost lies in $E_{\neg\varphi}^k$.

Given a subformula $\psi$ of $\varphi$, if we use $||E_\psi^k||$ to denote the max length of the Boolean formula describing that $\psi$ is initially satisfied upon a $k$-step unrolling, then it can be inductively computed as follows.

- $||E_\bot^k|| = ||E_\top^k|| = 0$. [4]
- $||E_p^k|| = 1$ for each $p \in \mathcal{P}$.
- $||E_{\varphi_1 \to \varphi_2}^k|| = ||E_{\varphi_1}^k|| + ||E_{\varphi_2}^k|| + 1$.
- $||E_{\mathbf{X}\psi}^k|| = ||E_{\mathbf{Y}\psi}^k|| = ||E_\psi^k||$.
- $||E_{\varphi_1 \mathbf{U} \varphi_2}^k|| = ||E_{\varphi_1 \mathbf{S} \varphi_2}^k|| = L(k) \times ||E_{\varphi_1}^k|| + k \times ||E_{\varphi_2}^k||$. [5]

---

[2] In [8], a "fixpoint"-based encoding is proposed, and we also attribute this technique to semantic encoding in this paper.

[3] Note that the part w.r.t. fairness constraints can be linearized.

[4] This is just for the case when $\bot$ or $\top$ appears as a subformula in the specification, and hence can be optimized; otherwise, we have $||E_\bot^k|| = ||E_\top^k|| = 1$.

[5] Note that this case does not imply that further blow-up would be caused with deeper nesting of temporal operators. For example, in [9], by introducing sharing propositions and reusing, it still leads to a linear encoding for the whole formula.

Here, $L(k)$ is some polynomial about $k$, related to the encoding approach. For example, with the technique proposed in [1, 8], we have $L(k) \in \mathcal{O}(k^2)$, whereas $L(k) \in \mathcal{O}(k)$ in [9]. This partly explains the reason that we tend to change temporal nestifications with Boolean combinations, as done in ($\mathrm{U}^{\mathrm{U}}[3 \to 2]$) etc.

Another feature affecting the cost is the number of propositions occurring in the encoding. If we denote by $var_k(\varphi)$ the set of additional propositions which only taking part in the encoding of $E^k_{\neg\varphi}$, then we have the following conclusions.

– For the techniques proposed in [1] and [4], we have $var_k(\varphi) = 0$. i.o.w., all propositions required in encoding $E^k_{\neg\varphi}$ can be shared with those for $E^k_M$.
– In term of the encoding presented in [9], we need to add $\mathcal{O}(k)$ new propositions to $var_k(\varphi)$ for each **U**-subformula and for each **S**-subformula.

**Theorem 3.** *Let "Cond $\triangleright\ \varphi \approx \psi$" be a reduction rule, then we have $||E^k_\psi|| \leq ||E^k_\varphi||$ and $|var_k(\psi)| \leq |var_k(\varphi)|$ in syntactic encodings.*

## 4 Experimental Results

We have integrated CEPRE as an upfront option in NUSMV.[6] We have conducted experiments upon both industrial benchmarks and random generated cases in terms of both BDD-based and bounded model checking (and the BMC encoding approach here we adapt is that proposed in [4], which is the current implementation of NUSMV).

We conduct the experiments under such platform: CPU - Intel Core Duo2 E4500 2.2GHz, Mem - 2G Bytes, OS - Ubuntu 10.04 Linux, Cudd -v2.4.1.1, Zchaff -v2007.3.12.

### 4.1 Experiments upon Industrial Benchmarks

The benchmark we choose in this paper is from [2], and most of them come from real hardware verification.

Table 1 provides experimental results for BDD-based LTL symbolic model checking. The field #Time is the summation of user time and system time, and the field #R.S. refers to the number of totally reachable states. For Table 1, we have the following remarks:

1. 8 out of 16 specifications could be reduced with CEPRE (and these specifications have been highlighted).
2. For the specifications that can be reduced, considerable improvements are made in allocating resources. The most significant case is Pit.g.ltl — with CEPRE, the number of BDD nodes are decreased to 12.5% of that without using CEPRE.

---

[6] The tool is available in `http://sourceforge.net/projects/nusmvwithcepre`, and all SMV manuscripts for experiments can be found in the folder of `/files/benchmark` and `/files/random` from that site.

|       |           | Without CePRe |           |           | With CePRe |           |           |
|-------|-----------|---------------|-----------|-----------|------------|-----------|-----------|
| Model | Spec.     | #BDD-Nodes    | #R.S.     | #Time (sec.) | #BDD-Nodes | #R.S.     | #Time (sec.) |
| srg5  | Ptimo.ltl | 7946          | 720       | 0.024     | 2751       | 720       | 0.016     |
|       | Pti.gnv.ltl | 29704       | 11460     | 0.058     | 5712       | 2880      | 0.012     |
|       | Pti.g.ltl | 64749         | 130048    | 0.048     | 8119       | 32768     | 0.016     |
| abp4  | P2false.ltl | 99577       | 559104    | 0.200     | 99625      | 559104    | 0.202     |
|       | P2true.ltl | 61209        | 904384    | 0.066     | 56494      | 419296    | 0.064     |
|       | Pold.ltl  | 52301         | 353536    | 0.060     | 52349      | 353536    | 0.064     |
|       | Ptimo.ltl | 78098         | 219616    | 0.080     | 78146      | 219616    | 0.088     |
|       | Pti.g.ltl | 8385          | 200704    | 0.060     | 8433       | 200704    | 0.062     |
| dme3  | P0.ltl    | 889773        | 35964     | 5.756     | 527983     | 26316     | 5.096     |
|       | P1.ltl    | 455148        | 8775      | 0.460     | 409432     | 5505      | 0.374     |
| dme5  | Mdl.ltl   | 793942        | 8.64316e+06 | 167.346 | 814494     | 3.2097e+06 | 114.599  |
|       | Wat.ltl   | 412867        | 1.79217e+07 | 302.005 | 967033     | 1.12567e+07 | 286.850 |
|       | Ptimo.neg | 508036        | 1.26202e+06 | 3.260   | 508081     | 1.26202e+06 | 3.280   |
| msi_w-trans | Sched.ltl | 2275558  | 7.31055e+07 | 6.612   | 2275655    | 7.31055e+07 | 6.632   |
|       | Safety.ltl | 1213308      | 3.6528e+07 | 7.568    | 1213460    | 3.6528e+07 | 7.644    |
|       | Seq.ltl   | 1921973       | 3.5946e+07 | 93.570   | 1702585    | 1.7973e+07 | 94.085   |

**Table 1.** Comparative results of BDD-based MC with/without CePRe.

3. Something noteworthy we do not provide in the table is that: if a violated LTL specification can be reduced, the newly generated counterexample is usually shorter than that of before. Among 8 specifications that can be reduced, counterexample-lengths of Pti.nuv.ltl, Pit.g.ltl, P0.ltl and Seq.ltl are respectively shortened to 15, 10 and 194, opposing to the original values 16, 12 and 217. Meanwhile, counterexample-lengths of others are kept unchanged.

Table 2 yields the experimental results for BMC-based model checking, and we here give some comments on that.

1. With NuSMV, one need to preset a max-bound when doing bounded model checking. The column #Max-bound gives such values — a "star mark" means that this bound does not reach the completeness threshold. The field #N.O.C. designates the number of clauses generated during model checking.
2. From the table, we can see that without CePRe the specification Pti.gnv.ltl generates 2101 clauses when a counterexample is detected, in contrast, it only produces 299 clauses if CePRe is switched on.
3. Another impressive comparison is for P0.ltl upon dme3: If we don't do any reduction, the SAT-solver reports a SEGMENTATION FAULT at Step 35. In contrast, using CePRe, a counterexample could be found at Step 62.
4. Since the encoding approach we adapt is taken from [4], propositions used in the encoding are only determined by the model and the bound, thus the number of required propositions does not change. For this reason, the corresponding experimental results on proposition numbers are not provided.

| Model | Spec. | Without CePRe #N.O.C. | Without CePRe #Time (sec.) | With CePRe #N.O.C. | With CePRe #Time (sec.) | #Max-bound |
|---|---|---|---|---|---|---|
| srg5 | Ptimo.ltl | 272567 | 67.391 | 1371 | 0.143 | 20 |
|  | Pti.gnv.ltl | 2101 | 0.116 | 299 | 0.024 | 6 |
|  | Pti.g.ltl | 21 | 0.016 | 21 | 0.016 | 1 |
| abp4 | P2false.ltl | 7532 | 3.972 | 7532 | 3.972 | 17 |
|  | P2true.ltl | 12639 | 8.145 | 9369 | 7.753 | 20★ |
|  | Pold.ltl | 7499 | 9.087 | 7499 | 9.488 | 20★ |
|  | Ptimo.ltl | 6332 | 2.500 | 6332 | 2.512 | 16 |
|  | Pti.g.ltl | 11952 | 0.841 | 11952 | 0.976 | 20★ |
| dme3 | P0.ltl | – | – | 35102 | 524.207 | 62 |
|  | P1.ltl | 216 | 0.036 | 167 | 0.048 | 1 |
| dme5 | Mdl.ltl | 90 | 0.044 | 90 | 0.048 | 0 |
|  | Wat.ltl | 367 | 0.048 | 274 | 0.052 | 1 |
|  | Ptimo.neg | 367 | 0.050 | 277 | 0.058 | 1 |
| msi_w-trans | Sched.ltl | 14235 | 1.076 | 14235 | 1.078 | 20★ |
|  | Safety.ltl | 12439 | 8.441 | 12439 | 8.448 | 20★ |
|  | Seq.ltl | 1907 | 0.064 | 81 | 0.052 | 3 |

**Table 2.** Experimental results of BMC-based MC with/without CePRe.

It should be pointed that both model-independent and model-dependent rules contribute to the reductions. For example, for the model srg5 and specification Pti.g.ltl, the rules (FS) and (S) are applied; meanwhile, for the model msi_wtrans and the specification Seq.ltl, the rule ($U^U[\neg 2 \to 3]$) takes part in the reduction.

### 4.2 Experiments w.r.t. Random Models and Specifications

We have also performed experiments upon randomly generated models and specifications with the tool LBTT [13] and with the methodology suggested in [2].

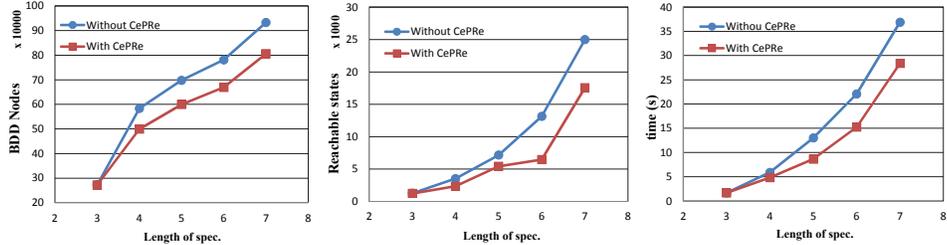

**Fig. 9.** Experimental results on BDD-based model checking for random cases.

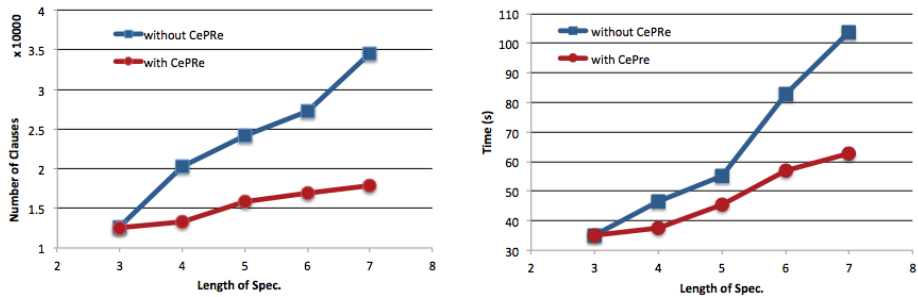

**Fig. 10.** Experimental results on bounded model checking for random cases.

For each $3 \leq \ell \leq 7$, we randomly generate 40 specifications having length $\ell$. Subsequently, for each specification, we generate two models respectively for the BDD-based model checking and for BMC. Hence, we totally have 200 specifications and 400 models.

For the BDD-based model checking, we give the comparative results on 1) the scale of BDD-nodes, 2) the number of reachable-states, and 3) the time consumed, as shown in Figure 9. For BMC, we have set the max-bound to 20 and we have compared 1) the number of clauses, and 2) the executing time, as shown in Figure 10. Each value here we provide is the average of the 40 executions.

For the BDD-based model checking, there are 123 (out of 200) specifications can be reduced; whereas for BMC, the number of specifications that can be reduced is 118.

## 5  Concluding Remarks

In this paper, we present a new technique to reduce LTL specifications' complexity towards symbolic model checking, namely, CePRe. The novelty in this technique is that the formula being reduced need not to be logically equivalent with the one after reduction, but just need to preserve the counterexample set. Moreover, the condition enabling such a reduction can be usually detected with lightweight approaches, such as SAT-solving. Hence, this technique could leverage the power of SAT-solvers.

The central part of CePRe is a set of reduction rules, and soundness of these reduction rules are fairly easy to check. For the model dependent rules, additional conditions mainly concern invariants and transitions only, and we do not make a sufficient use of other features, such as fairness. In this paper, we just consider combinations of two temporal operators as many as possible, indeed, there might be other possible reduction schemas we are not aware. Indeed, in this paper, we tentatively to provide such a framework, and one can extend it to model checking of other logics.

From the experimental results, we can see that, in a statistical perspective, we can gain a better performance and lower overhead with CePRe.